\newcommand{\vac}[1]{\langle #1 \rangle}
\newcommand{\eqs}[1]{\begin{equation} #1 \end{equation}}
\newcommand{\beqa}{\begin{eqnarray}}
\newcommand{\eeqa}{\end{eqnarray}}
\newcommand{\beqn}{\begin{eqnarray*}}
\newcommand{\eeqn}{\end{eqnarray*}}

\renewcommand{\theequation}{\thesection.\arabic{equation}}

\newcommand{\cl}{\centerline}

\documentstyle[12pt]{article}
\begin{document}
\hfill{UTHEP-241}\par
\hfill{July 1992}\par
\setlength{\textwidth}{5.0in}
\setlength{\textheight}{8.0in}
\setlength{\parskip}{0.1in}
\setlength{\baselineskip}{15.5pt}
\pagestyle{empty}
\par\bigskip
\par\bigskip
\cl{\large{\bf Fermion Interactions in the Wilson Yukawa Approach}} \par
\cl{\large{\bf for Lattice Chiral Gauge Theories }}
\par\bigskip
\par\bigskip
\cl{{\bf Sinya AOKI}\footnote{saoki@ph.tsukuba.ac.jp}}
\vskip .1in
\cl{ Institute of Physics,}
\cl{ University of Tsukuba,}
\cl{Tsukuba, Ibaraki 305, Japan}
\par\bigskip
\par\bigskip
%\cl{\today }\par
\flushbottom
\par\bigskip
\par\bigskip
\cl{{\bf Abstract}}\par
We consider fermion-gauge couplings in the Wilson-Yukawa approach for lattice
chiral gauge theories.
At the leading order of a fermionic hopping parameter expansion
we find that the fermion-gauge coupling has a chiral and tree-like
structure. We argue that this fermion-gauge coupling remains non-zero in the
continuum limit taken in the Higgs phase.
Possible fermion-scalar couplings in this approach are considered.
We also evaluate the fermion interaction with an external gauge field in
the slightly modified model and show that it has a chiral structure in general.
\par\bigskip
\vfill
\noindent
\vfill

\section{Introduction} \par
\setcounter{equation}{0}
\pagestyle{plain}
\setcounter{page}{1}
\renewcommand{\theequation}{\arabic{section}.\arabic{equation}}

    Construction of chiral gauge theories on the lattice is an
important task.  The prime motivation for this effort is that one
would like to have full understanding of the standard SU(2) $\times$
U(1)$_Y$ electroweak theory, which subsumes, but is not restricted to,
perturbation expansions.  Indeed, although the standard
electroweak model has been quite successful
phenomenologically, it leaves a number of fundamental questions unanswered,
including the nature of the symmetry breaking and, in particular, why the
chiral gauge symmetry SU(2) $\times$ U(1)$_Y$ is realized in a "spontaneously
broken" mode, whereas the vectorial gauge invariances of SU(3)$_{color}$
and U(1)$_{e.m.}$ are
realized symmetrically (in confined and deconfined modes,
respectively).  The conventional Higgs mechanism does not explain this
since, among other things,
no {\it ab initio} reason is given for choosing the symmetry-breaking sign
of the coefficient of $\phi^{\dagger}\phi$ in ${\cal L}_{ew}$.
Partially in response to this inadequacy and
the problems of the quadratic mass shift and necessity of fine-tuning
in the Higgs sector, appealing alternative mechanisms have been proposed
involving dynamical symmetry breaking without any fundamental scalars.
However, one
knows very little about the properties of such pure chiral
gauge theories.  A better understanding of these
might well yield a deeper insight into how the electroweak symmetry
is realized in nature.
Moreover, since the top quark is now known to have a mass greater than the $W$
boson, the corresponding physical
Yukawa coupling in the standard model is of order unity, so that in
studying the physics of the top quark, it is desirable to use a method
which is capable of nonperturbative calculations.

 The lattice approach provides a potentially
useful tool for all of these purposes since
it deals with the full path integral and is not limited to perturbative
expansions in couplings.  However, there has been a serious difficulty in
applying this method: because of fermion doubling, each
lattice fermion field yields $2^d=16$ fermion modes, half of one chirality
and half of the other, so that the lattice
theory is nonchiral (e.g. Ref. \cite{NN}).
There are several lattice formulations of the chiral gauge
theories. Among these formulations
the Wilson-Yukawa approach\cite{swift,smit1,aoki1,FK1}
has been extensively studied during the past few
years\cite{ALX,ALSS,ALS,Aachen}.
It is now established that the decoupling of fermion doublers
can be made in the strong Wilson-Yukawa coupling region.
Therefore it seems that this approach successfully describes
lattice chiral gauge theories.
However
base on a recent numerical study\cite{BDS} which suggests that only gauge
singlet fermions become light in the symmetric phase of the fermion-scalar
system in this approach,
it is claimed\cite{GPS} that the gauge-fermion interaction
vanishes in the continuum limit in the order of $a^2$ where $a$ is a
lattice spacing, and therefore that the Wilson-Yukawa approach fails to
describe
the chiral gauge interaction.
In this paper we consider this problem in detail to see whether
the Wilson-Yukawa approach indeed fails or not.
We calculate the gauge-fermion coupling using a fermionic hopping parameter
expansion and find that
it is likely that the fermion-gauge coupling in the {\it Higgs} phase remains
non-zero in the continuum limit.
The Wilson-Yukawa approach is still alive.

In the next section, we define a simple model in this approach
where a right-handed part of the fermion is a gauge singlet. We briefly
summarize how the fermion doublers are decoupled in the continuum limit.
In sect. III we calculate the gauge fermion interaction\cite{aoki3}
using the hopping parameter expansion and
find that the fermion-gauge coupling has a chiral and tree-like
structure. We argue that this fermion-gauge coupling remains non-zero in the
continuum limit taken in the Higgs phase.
There we discuss a possible explanation
why the gauge interaction in the standard model is perturbative.
A Yukawa coupling is also considered.
In sect. IV we calculate a charged fermion interaction with an external gauge
field in the modified model given in ref.\cite{GPS} where it is
claimed that this coupling becomes vector-like.  We show that this coupling
appears to be chiral in general.
A conclusion is given in sect. V and
a more complicated model where both left- and right- handed fermions are
gauge non-singlet are considered.

\section{The Model and the Decoupling of Fermion Doublers} \par
\setcounter{equation}{0}
\renewcommand{\theequation}{\arabic{section}.\arabic{equation}}

We define a lattice chiral gauge theory with a right-handed fermion being
gauge singlet in the Wilson-Yukawa approach. The action is given by
\eqs{
S = S_G + S_H + S_F + S_Y + S_{WY}
}
where $ S_G$ is the action for the gauge link variable $U_{n,\mu} \in U(1)$
or $SU(2)$,
\eqs{
S_H = \beta_h \sum_{n,\mu} r_n r_{n+\mu} Re [ d_N {\rm Tr} ( g_n U_{n,\mu}
g_{n+\mu}^\dagger) ] - \lambda \sum_n [(r_n^2 - 1)^2 + r_n^2 ]
}
is the action for a Higgs field $\phi_n = r_n g_n$ with $r_n \in R$ defined
by $r_n^2 =  \phi_n^\dagger \phi_n $, $g_n \in U(1)$ or $SU(2)$
satisfying $g_n^\dagger g_n = 1$, and $d_N = 1 $ for the $U(1)$ case
and $d_N = 1/2$ for the $SU(2)$ case.
$S_F + S_Y +S_{WY}$ is the fermionic part of the action where
\eqs{
S_F = \frac{1}{2}\sum_{n,\mu}\bar\psi_n [ ( U_{n,\mu} P_L + P_R )
\psi_{n+\mu} - ( U_{n,-\mu} P_L + P_R ) \psi_{n-\mu} ]
}
is the naive fermion action with $P_{L/R} =\displaystyle
\frac{1\pm\gamma_5}{2}$,
\eqs{
S_Y = y \sum_n r_n \bar\psi_n ( g_n P_L + g_n^\dagger P_R )\psi_n
}
is the Yukawa interaction with a Yukawa coupling $y$, and
\beqa
S_{WY} & = -\frac{r}{2}\sum_{n,\mu}
\bar\psi_n [ P_L (g_{n+\mu} \psi_{n+\mu}+g_{n-\mu} \psi_{n-\mu}
- 2g_{n} \psi_{n} ) \nonumber \\
& \qquad +P_Rg_n^\dagger (\psi_{n+\mu}+\psi_{n-\mu}-2\psi_n) ]
\eeqa
is the Wilson-Yukawa term with the Wilson-Yukawa coupling $r$.

There are few remarks for the above action.
\begin{enumerate}
\item
Only $g$ field instead of $\phi$
appears in the Wilson-Yukawa term.  If $\phi$ is introduced
in the Wilson-Yukawa term, we have to tune the Wilson parameter $r$
in order to decouple the fermion doublers in the continuum limit.
Although this is possible, it is not so easy in this case
to formulate the Hopping parameter expansion
Therefore we do not introduce $\phi$ in the Wilson-Yukawa term in this paper.
\item
The lattice Higgs action $S_H$ is obtained from
the usual continuum (cn) parametrization\cite{shrock1}
\eqs{
S_{H,cn} = -\int d^d x [ | D_\mu \varphi_{cn} |^2
- \lambda_{cn} (|\varphi_{cn}|^2 - v_{cn}^2 )^2 ]
}
under the substitutions
\beqn
& \varphi_{cn} a = \sqrt{\displaystyle \frac{\beta_h}{2}} \phi  \\
& v_{cn}^2 a^2 =\displaystyle \frac{\beta_h}{2}[1 +
\displaystyle \frac{( \beta_h d - 1)}{2\lambda} ] \\
& \lambda_{cn} =\displaystyle \frac{4}{\beta_h^2} \lambda
\eeqn
where $d$ is the space-time dimension and $a$ is the lattice spacing.
The $\lambda \rightarrow \infty$ limit is usually taken in the
Wilson-Yukawa approach so that the Higgs field becomes non-linear $\sigma$
model ( : $\forall n$, $ r_n = 1$ ). The shift symmetry of the above
action\cite{GP}
prohibits Yukawa interactions between the gauge singlet fermion and the
non-linear Higgs field in this limit.
In this paper we use the linear Higgs field instead
in order to allow the Yukawa interaction. See also ref.\cite{AK1}.
\item
It is also noted that the vacuum expectation value of $r_n$ is not equal to
the
vacuum expectation valu of $\phi_n$ : $ \vac{\phi_n} = \vac{ r_n g_n}
\not= \vac{r_n}$. This $\vac{r_n}$ is always non-zero and
\eqs{
\lim_{\lambda\rightarrow\infty} \vac{r_n} = 1  .
}
We therefore define $\rho_n \equiv r_n - \vac{r_n} $ as the dynamical
variable.
\end{enumerate}

\bigskip

The gauge singlet fermion field $\chi$ is defined by
\eqs{
\chi_n = \frac{1}{\sqrt{2K}} (g_n P_L + P_R)\psi_n, \qquad
\bar\chi_n = \frac{1}{\sqrt{2K}} \bar\psi_n(g^\dagger_n P_R + P_L)
}
where $K =\displaystyle \frac{1}{2(\vac{r_n}y+dr)}$ is the hopping parameter.
Since the gauge singlet fermion field is expected to appear as a particle
in the strong Wilson-Yukawa coupling region, we rewrite the fermionic part
of the action in terms of $\chi$ and obtain
\eqs{
S_F+S_{WY}+S_Y =
\sum_{n,\mu} \bar\chi_n [ (1+2K\rho_n)\chi_n - K
( T_{n,\mu} \chi_{n+\mu} + T_{n-\mu}\chi_{n-\mu} ) ]
\label{HPE1} }
where
\beqn
T_{n,\mu} & = r-\gamma_\mu ( U^g_{n,\mu} P_L + P_R ) \\
T_{n,-\mu} & = r+\gamma_\mu ( U^g_{n,-\mu} P_L + P_R )
\eeqn
and $ U^g_{n,\mu} \equiv g_n U_{n,\mu} g_{n+\mu}^\dagger$ is the
gauge singlet link variable.
If $K$ is small, we can expand fermionic quantities in terms of $K$
and we call this expansion the hopping parameter expansion( HPE
)\cite{ALX,ALS}.
It is noted that the large $r$ expansion\cite{GPR} is also applicable to
our simple
action. We check that both expansions give the same results for the quantities
we calculate in this papaer.

In order to calculate the fermion propagator, we write down a recursion
relation for the fermion:
\eqs{
\vac{ \chi_n \bar\chi_m}
= -\delta_{nm}
- 2K \delta_{nm} \vac{\rho_n \chi_{n}\bar\chi_m }
+K \sum_{\pm\mu} \vac{ T_{n,\mu} \chi_{n+\mu}\bar\chi_m } .
}
Using the truncation of the recursion relation and the fact that
$\vac{\rho_n} = 0$ we obtain
\eqs{
\vac{ \chi_n \bar\chi_m}
= -\delta_{nm}
+K \sum_{\pm\mu} \vac{ T_{n,\mu}} \vac{\chi_{n+\mu}\bar\chi_m }  .
}
The fermion propagator is calculated as
\eqs{
G_F(n-m) \equiv \vac{\chi_n\bar\chi_m}
= - \left[ 1 - K \sum_{\pm\mu} T_\mu \nabla_\mu \right]_{nm}^{-1}
}
where
$ T_{\pm\mu} = r \mp \gamma_\mu ( L P_L + P_R )$,
$L = \vac{U^g_{n,\mu}}$ and $[\nabla_{\pm}]_{nm} =\delta_{m,n\pm\mu}$.
In the momentum space it becomes
\eqs{
G_F(p)^{-1} = - 2K[ \sum_\mu \gamma_\mu \sin p_\mu (L P_L + P_R) + y \vac{r_n}
+ r\sum_\mu (1-\cos p_\mu) ]  .
}
The physical fermion mass in lattice unit is given by
\eqs{
m_{phy} a = y \vac{r_n} Z_F^{1/2}
}
while the mass of the fermion doubler becomes
\eqs{
m_d a = (y \vac{r_n} + 2 r l ) Z_F^{1/2}
}
where $Z_F = \displaystyle \frac{1}{L}$ and $l$ denotes the number of $\pi$'s
for the doubler momenta. (We call such doublers as $l$-th doublers.)
{}From the above formula, it is easy to see the following
results\cite{ALX,ALSS,ALS,Aachen}.
\begin{enumerate}
\item We have to tune the Yukawa coupling $y$ such that
$\lim_{a\rightarrow 0} y = 0$ in order to get
the finite $m_{phy}$ in the continuum limit since
$\lim_{a\rightarrow 0} \vac{r_n}\cdot Z_F^{1/2} \not= 0$ .
This also shows that the fermion mass stays non-zero even in the
symmetric phase of the scalar model where the vacuum expectation value
of the Higgs field, $v_R =\vac{\phi^R_n}$, vanishes.
Therefore, the perturbative relation $m_{phy} = Y_R v_R$
does not hold in the strong Wilson-Yukawa coupling region, where
$Y_R$ is the renormalized Yukawa coupling.
\item With this tuning of $y$, the fermion doublers are decoupled
in the continuum limit since $\lim_{a\rightarrow 0} m_d a =
2l r\times  Z_F \not = 0$.
\end{enumerate}

In the next section we calculate fermion couplings\cite{aoki3} using the HPE.

\section{Fermion Couplings}\par
\setcounter{equation}{0}
\renewcommand{\theequation}{\arabic{section}.\arabic{equation}}

In the action (\ref{HPE1})
the gauge singlet fermions $\chi$,
$\bar\chi$ couple to the singlet link variable $U^g_{n,\mu}$.
Since naively
\eqs{
U^g_{n,\mu} \simeq U^T_{n,\mu} {\rm (transverse\quad mode) } \ +
\ g_n g_{n+\mu} {\rm  (longitudinal\quad mode) } ,
}
the action describes an interaction between the singlet fermion and
a {\it massive} vector field. This is the type of interactions needed for the
electro-weak interaction.
We have to study more carefully, however,
whether this is true or not
since the quantum fluctuation of $g$ field may invalidate the naive
expectation.
In the scaling region of the lattice gauge-Higgs system,
the operator $U^g_{n,\mu}$ is expected to interpolate the Higgs fields as well
as the gauge field such that
\eqs{
U_{n,\mu}^g = < U_{n,\mu}^g > +  H(n) + i  A_\mu^a(n) T^a+ \cdots
\label{Higgs1} }
where  $H$ is the spin-0 interpolation field for the Higgs field, $A_\mu$
is the spin-1 interpolation field for the gauge field, and $T^a$
which satisfies $ Tr (T^a T^b) = \delta_{ab}$ is the
generator of the gauge group in the representation of the left-handed fermion.
  Other excited states, which are denoted by $\cdots$ above,  may exist,
but we neglect them hereafter.
Since these operators are dimensionless on the lattice, the two points
functions of these operators in the scaling region become
\eqs{
G^H (p)  = \sum_n < H(0) H(n) > e^{ipa\cdot n} \nonumber
\simeq \displaystyle \frac{Z_H}{p^2 a^2 + m_H^2 a^2}
}
\eqs{
\delta_{ab} G^A_{\mu\nu} (p)  = \sum_n < A^a_\mu(0) A^b_\nu(n) >
e^{ipa\cdot n} \nonumber
\quad \simeq \displaystyle \delta_{ab}
\frac{Z_A}{p^2 a^2 + m_G^2 a^2} \left
(\delta_{\mu\nu}+\frac{p_\mu p_\nu}{m_G^2} \right)
}
in the momentum space, where $m_G$( $m_H$ ) is the dimensionful mass of gauge
(Higgs) field. The residue $Z_G$( $Z_H$ ) becomes the wave-function
renormalization constant for the operator $A_\mu$( $H$ ) and it gives the
overlapping between this operator and the possible asymptotic state for the
gauge( Higgs ) field. If $Z_G$( $Z_H$ ) remains non-zero in the continuum
limit, the asymptotic gauge( Higgs ) field exists in this theory and
it couples to $A_\mu$ ( $H$ ).
This also shows how a linear Higgs field $H$ can appear from the
non-linear field $g$. We hope that the appearance of the linear Higgs field
 leads to a renormalizable
theory in the scaling region even if the original lattice action is not
perturbatively renormalizable due to the non-linearity of $g$ in
the Wilson-Yukawa term.

Since the operator $\rho_n$ can also interpolate the Higgs fields, we
can define
\eqs{
G^\rho (p)  = \sum_n < \rho_0 \rho_n > e^{ipa\cdot n}
\simeq \displaystyle \frac{Z_\rho}{p^2 a^2 + m_H^2 a^2} .
\label{Higgs2}}
{}From eq. \ref{Higgs1} and eq. \ref{Higgs2}, we also obtain
\eqs{
G^{H \rho} (p)  = \sum_n < H(n) \cdot\rho_n > e^{ipa\cdot n}
\simeq \displaystyle \frac{\sqrt{Z_H Z_\rho}}{p^2 a^2 + m_H^2 a^2} .
}

Now we demonstrate how the fermion coupling with the gauge( Higgs ) field is
calculated in the hopping parameter expansion\cite{aoki3}.
First we introduce source terms
for the gauge field and the Higgs field in the action:
\eqs{
S + \sum_n J^H_n H(n) + \sum_{n,\mu}J^a_{n,\mu} A_\mu^a(n) .
}

Next
we write down the previous recursion relation for the fermion propagator
in the presence of the source terms. Using the same truncation of
the previous calculation we obtain
\eqs{
\vac{ \chi_n \bar\chi_m}_J  = -\delta_{nm}
-2K\delta_{nm} \vac{\rho_n}_J\vac{\chi_{n}\bar\chi_m}_J
\quad + K \sum_{\pm\mu}  \vac{ T_{n,\mu} }_J \vac{\chi_{n+\mu}\bar\chi_m}_J
}
where $\vac{\ \cdot \ }_J$ is the vacuum expectation value in the presence
of the
source $J$'s. Taking the derivative with respect to the sources and putting
them zero we obtain the following three points functions
\beqa
\vac{\chi_n \bar\chi_m A^a_\mu (l)} & = i K\sum_{s,\nu} G_F(n-s)\gamma_\nu T^a
[ G_F(s+\nu-m)G^A_{\nu\mu}(s+\nu/2-l)  \nonumber \\
& \  +G_F(s-\nu-m)G^A_{\nu\mu}(s-\nu/2-l)]
\eeqa
\beqa
\vac{\chi_n \bar\chi_m H(l) } & = K \sum_{s,\nu} G_F(n-s)\gamma_\nu
[ G_F(s+\nu-m)G^H(s+\nu/2-l) \nonumber \\
& \quad -G_F(s-\nu-m)G^H(s-\nu/2-l)]  \nonumber \\
& \quad + 2K y \sum_s G_F(n-s) G_{\rho H}(s-l) G_F(s-m)  .
\eeqa
\ From the above expression we obtain the renormalized fermion-gauge vertex
denoted as $ \Gamma_\mu^a(p,q)$ in momentum space;
\eqs{
\Gamma_\mu^a (p,q) = i \gamma_\mu P_L T^a \cos\frac{(p+q)_\mu a}{2} Z_G^{1/2}
Z_F
}
where  $p$ and $q$ are fermion momenta with $p-q$ being the
momentum transfer to the
gauge field.
This vertex has the correct chiral structure ( $\gamma_\mu P_L$ )
and gauge structure( $T^a$ ).  In the continuum limit the coupling constant
$g_F^r$ becomes
\eqs{
g_F^r = \lim_{a\rightarrow 0}\frac{Z_G^{1/2}}{L}  .
}
It is generally believed that
the continuum limit of the gauge-Higgs system can be taken
at ( $\beta_g$ , $\beta_h$ ) $=$ ( $\infty$ , $\beta_h^c$ ), where
$\beta_g$ is the inverse gauge coupling, and $\beta_c$ is the second order
phase transition point of the Higgs system without gauge field.
At this continuum limit, it is easy to see that $ 0 < L < 1 $.
If the continuum limit taken in the broken phase gives a non-zero $Z_G$;
\eqs{
\lim_{\beta_g \rightarrow \infty, \beta_h \searrow \beta_h^c}
Z_G \not= 0 ,
}
the fermion-gauge interaction remains non-zero in the continuum limit.
If this is the case, the Wilson-Yukawa approach
can describes the fermion interaction with the massive gauge
field. Therefore it is very crucial and important
for the Wilson-Yukawa approach to investigate the behaviour of $Z_G$ near
the continuum limit.  For example, the  perturbative expansion in
the unitary gauge ($\forall g_n=1$)
gives $ L =1 + O( g_r^2 )$ and $Z_G = g_r^2 + O( g_r^4 )$\footnote{
It is noted that the lattice gauge field $A_\mu$ is related to the
continuum gauge field $A_\mu^{con}$ such that
$ A_\mu = a g_r A_\mu^{con}$.
}
where
$g_r$ is the renormalized coupling of this perturbative expansion, and
the formula for $g_F^r$ give the consistent relation $ g_F^r = g_r$.
The perturbative calculation, however, may not be reliable due to the large
quantum fluctuation of $g$ field near the scaling region even in the
broken phase.  In order to obtain the behaviour of $Z_G$ more reliably
we have to calculate it numerically. This numerical calculation should be
done.

Assuming that the Wilson-Yukawa approach gives the non-zero
coupling, we may explain why the electroweak interaction is well described
by the perturbative expansion of the coupling constant. The argument is the
following.  Since only gauge singlet fields such that $\chi$, $\bar\chi$ and
$U^g$ can appear as the asymptotic states in
the strong Wilson-Yukawa coupling region where the fermion doubler
is removed, interactions among these neutral fields is expected to be weak.
Indeed the hopping parameter expansion which is valid in the strong
Wilson-Yukawa region show that the fermion-gauge vertex is dominated by the
tree-like diagram. Furthermore higher order corrections to the vertex
in the hopping parameter expansion correspond
to those of the perturbative expansion of the
gauge coupling constant. This is the reason why the perturbative theory
successfully describes the fermion-gauge interaction
in the region where the hopping parameter expansion is reliable.

Recently it is claimed that the fermion-gauge coupling vanishes as
$O(a^2)$ in the continuum limit
taken in the symmetric phase. This is equivalent
to the behaviour $ Z_G = O(a^4)$ near the scaling region.
This behaviour may be true in the symmetric phase,
since the symmetric phase may be similar to the confining theory and
only glue balls, not the gauge field, can appear as asymptotic sates
in the confining theory. The Wilson-Yukawa approach may fail
to describe the symmetric phase of chiral gauge theories.
Even if this is the case, it does not mean that the Wilson-Yukawa approach
fails to describe the standard model in the broken phase.
Since we expect that the massive gauge field appear in the broken phase,
it is likely that $\lim_{a\rightarrow 0} Z_G \not= 0$ in the broken phase.

Finally the renormalized fermion-Higgs coupling
$ \Gamma_H(p,q)$ is calculated as
\eqs{
\Gamma_H (p,q) = i\sum_\mu\gamma_\mu P_L  \sin\frac{(p+q)a}{2}
 Z_H^{1/2} Z_F  + y Z_\rho^{1/2}Z_F
}
The first term, which has been calculated for the case of the
non-linear Higgs field,
is the derivative coupling with dimension 5
and this coupling vanishes as $O(a)$ in the continuum limit\cite{aoki3}.
The second term is the ordinary Yukawa coupling which becomes
\eqs{
y \lim_{a\rightarrow 0}  Z_\rho^{1/2} Z_F
}
in the continuum limit\cite{AK1}. From the triviality of the
Yukawa interaction,
however, we expect that the Yukawa coupling
logarithmically goes to zero in the continuum limit.
Therefore $Z_\rho^{1/2} Z_F$ should logarithmically go to zero according
to the prediction of the renormalization group equation.

\section{Charged Fermion Couplings with the External Gauge Field} \par
\setcounter{equation}{0}
\renewcommand{\theequation}{\arabic{section}.\arabic{equation}}

In this section we consider fermion couplings with the external
gauge field in the slightly modified Wilson-Yukawa term\cite{GPS}:
\beqa
S_{WY} & = -\frac{r}{2} \bar\psi_n [ P_L g_n ( U_{n,\mu}\psi_{n+\mu}
+U_{n,-\mu}\psi_{n-\mu} - 2 \psi_n ) \nonumber \\
& \qquad +P_R ( U_{n,\mu}g_{n+\mu}^\dagger
\psi_{n+\mu}+U_{n,-\mu}g_{n-\mu}^\dagger\psi_{n-\mu}
- 2 g_{n}^\dagger\psi_n )] .
\eeqa
In terms of the vectorially charged
fermion field $ C_n$ defined as
\eqs{
C_n = \frac{1}{\sqrt{2K}} ( P_L + g_n^\dagger P_R)\psi_n = g_n^\dagger N_n
, \quad
\bar C_n = \frac{1}{\sqrt{2K}} \bar\psi_n( P_R + g_nP_L) = \bar N_ng_n ,
}
the fermionic part of the action becomes
\eqs{
S_F + S_Y + S_{WY} =
\sum_n \bar C_n [ C_n - K \sum_{\mu} (T^c_{n,\mu} C_{n+\mu} +
T^c_{n,-\mu} C_{n-\mu} )]
}
where
\beqa
T^c_{n,\mu} & = U_{n,\mu} (r -\gamma_\mu P_L) -
R_{n,\mu} \gamma_\mu P_R  \nonumber \\
T^c_{n,-\mu} & = U_{n,-\mu} (r +\gamma_\mu P_L) +
R_{n,-\mu} \gamma_\mu P_R
\eeqa
and $ R_{n,\mu} = g_n^\dagger g_{n+\mu}$.  For simplicity we take the
$\lambda\rightarrow 0$ limit here( $\vac{r_n} =1$ and $\rho_n =0$ ).
It is noted that the charged fermion propagator in the global limit
( : $ \forall U_{n,\mu} = 1$ ) of this model is
reliably calculated by the hopping parameter expansion,
and  we obtain
\eqs{
G_F^c(p) =\sum_n \vac{ C_0 \bar C_n} e^{ip\cdot n} =
-(2K)^{-1} [ i\sum_\mu \gamma_\mu \sin p_\mu ( P_L +Z_R^{-1} P_R) +
M(p) ]^{-1}
}
where $Z_R^{-1} = R \equiv \vac{ R_{n,\mu}}$.
This is the reason why we use the modified action instead of the one in the
previous two sections.

Introducing the external gauge field $V^a_\mu(n) T^a$ such that
\eqs{
U_{n,\mu} = \exp [ i V^a_\mu(n) T^a ]
}
we define the current $J_\mu^a$ to which the external gauge field $V_\mu^a$
couples:
\eqs{
J_\mu^a(n) = \left[ \frac{\delta S}{i\delta V_\mu^a(n)} \right]_{V_\mu =0} .
}
The gauge current becomes
\beqa
J_\mu^a(n) & =
\beta_h d_N {\rm tr}( g_n T^a g_{n+\mu}^\dagger - g_{n+\mu}T^a g_n^\dagger )
\nonumber \\
& \quad
+ K [ \bar C_n T^a(\gamma_\mu P_L-r)C_{n+\mu}+\bar C_{n+\mu}T^a (\gamma_\mu P_L
+r) C_n ]  .
\eeqa
We calculate the 3 points function
using the hopping parameter expansion. By a method similar to that
in the previous sections, we obtain
\beqa
 \vac{ C_n^i \bar C_m^j J_\mu^a (l) } = K (T^a)^{ij}[
d_N \beta_h \sum_{s,\nu} G_F(n-s)
%\times
\{ G_{\mu\nu}^B(l-s) \gamma_\nu P_R \nonumber \\
 \times G_F(s+\nu-m)
+G_{\mu\nu}^B(l-s+\nu ) \gamma_\nu P_R G_F(s-\nu-m) \}
 + G_F(n-l) \nonumber \\
 \times (\gamma_\mu P_L - r) G_F(l+\mu-m)
 +G_F(n-l-\mu )(\gamma_\mu P_L + r) G_F(l-m) ]
\eeqa
where
\eqs{
\delta_{ab}G_{\mu\nu}(l-s) = \vac{ B^a_\mu (l) B^b_\nu (s) }
}
and $B^a_\mu (n) =
Im [ {\rm tr}( g_{n+\mu}T^a g_n^\dagger ) ]$ .
 After the renormalization for the fermion field such that
\eqs{
C_n = \frac{1}{\sqrt{2K}}(Z_R^{1/2}P_R + P_L) C_n^r, \quad
\bar C_n = \frac{1}{\sqrt{2K}} \bar C_n^r(Z_R^{1/2}P_L + P_R)
}
and
\eqs{
G_F(n) = \frac{1}{2K}(Z_R^{1/2}P_R + P_L) G_F^r (n) (Z_R^{1/2}P_L + P_R)
}
(Here $r$ stands for the {\it renormalized} .),
we obtain
\beqa
& \vac{ C_n^r \bar C_m^r J_\mu^a (l) } =
\displaystyle \frac{T^a}{2} [
d_N \beta_h Z_R \sum_{s,\nu} G_F^r(n-s)
%\times
\{ G_{\mu\nu}^B(l-s) \gamma_\nu P_R
G^r_F(s+\nu-m)   \nonumber \\
& +G_{\mu\nu}^B(l-s+\nu ) \gamma_\nu P_R G^r_F(s-\nu-m) \}
+ G^r_F(n-l)
(\gamma_\mu P_L - Z_R^{1/2}r)  \nonumber \\
& \times G^r_F(l+\mu-m)
 + G^r_F(n-l-\mu )(\gamma_\mu P_L + Z_R^{1/2}r) G^r_F(l-m) ] .
\eeqa
This result shows that the charged fermion interaction with the external
gauge field is chirally asymetric: the left-handed fermion and the
right-handed fermion have differents couplings with the external gauge field.

In the limit that $\beta_h \rightarrow 0$ of the above result,
we can reproduce the result in ref.\cite{GPS}.  Since in this limit
\eqs{
G_{\mu\nu}(l-s)^B = \frac{1}{2}\delta_{\mu,\nu} \delta_{l,s}
}
and
$Z_R^{-1} \equiv R = \displaystyle \frac{d_N}{2} \beta_h +O(\beta_h^2)$,
we obtain
\beqa
\lim_{\beta_h\rightarrow 0}
\vac{ C_n^r \bar C_m^r J_\mu^a (l) } & =\displaystyle \frac{T^a}{2} [
+ G^r_F(n-l)(\gamma_\mu  - Z_R^{1/2}r) G^r_F(l+\mu-m)
\nonumber \\ & \qquad +
G^r_F(n-l-\mu )(\gamma_\mu + Z_R^{1/2}r) G^r_F(l-m) ]  .
\eeqa
This is the result of ref.\cite{GPS} .
Their claim that the charged fermion coupling with the external gauge field
becomes vector-like is only true in the special limit that $\beta_h
\rightarrow 0$ and the coupling in general is chirally asymetric.

If we include the effect of the fermion determinant by the HPE\cite{ALX,ALS},
we have to replace $\beta_h$ with $\beta_h^{eff}=\beta_h + 2^{d/2} K^2$
in  all results of this paper.

\vspace{1cm}
\section{Conclusion and Discussion} \par
\setcounter{equation}{0}
\renewcommand{\theequation}{\arabic{section}.\arabic{equation}}

In this paper we calculate the fermion couplings using the hopping parameter
expansion. We obtain the following results.

\begin{enumerate}
\item
The interaction between the gauge singlet fermion and the massive gauge field
has a chiral structure and it likely stays non-zero in the continuum limit
taken within the Higgs phase.  Since we can take the unitary gauge such that
$\forall g_n =1$ in the Higgs phase, there is no essential difference between
the singlet fermion and the charged fermion.
Therefore it is concluded that the Wilson-Yukawa approach can correctly
describe the fermion-gauge interaction of the standard model.
Since the fermion-gauge interaction in the HPE is dominated by the tree
diagram of the ordinary perturbative expansion of the coupling,
the success of the perturbative expansion in the standard model is understood
by the goodness of the HPE in the strong Wilson-Yukawa coupling egion.
\item
By introducing the linear part of the Higgs fields, $r_n$, in the action,
the non-zero Yukawa coupling is calculated in the Wilson-Yukawa approach.
We show that the Yukawa coupling at the leading order of the HPE
is identical to the perturbative Yukawa coupling of the standard model.
The Wilson-Yukawa approach can correctly
describe the Yukawa interaction of the standard model.
\item
In the slightly modified model where the charged fermion is reliably treated
by the HPE, the charged fermion interaction with the external gauge field is
calculated. We show that the interaction in general is chirally asymetric.
Therefore, it is not so obvious that the symmetric phase of this theory becomes
a vector-like theory as claimed in ref.\cite{GPS} .
\end{enumerate}

The above conclusion can easily be extended to the more complicated models
where both the left-handed part and the right-handed part of fermion are
gauge non-singlet.
(The hypercharge sector of the standard model is an example of such models.)
For example, the fermion-gauge coupling is calculated as follows.
The covariant derivative part of the fermion action is given by
\beqa
S_F & = \frac{1}{2}\sum_{n,\mu}\bar\psi_n [ ( D^L(U_{n,\mu}) P_L
+ D^R(U_{n,\mu})P_R )\psi_{n+\mu} \nonumber \\
& - ( D^L(U_{n,-\mu}) P_L + D^R(U_{n,-\mu})P_R ) \psi_{n-\mu} ]
\eeqa
where
$D^{L(R)}$ is the representation of the gauge field for the left(right)-handed
fermion.  By the method of sect. III, we obtain the following
renormalized gauge-fermion vertex in the continuum limit.
\eqs{
\Gamma_\mu^a(p,q) = i\gamma_\mu ( g_L^r D^L(T^a) P_L + g_R^r D^R(T^a) P_R)
}
where the renormalized couplings are given by
\eqs{
g_L^r = (Z_A^L)^{1/2} Z_F^L , \qquad g_R^r = (Z_A^R)^{1/2} Z_F^R  .
}
Here $ Z_F^L $ and $ Z_F^R $ are the wave function renormalizations for the
fermion fields and the HPE gives $ Z_F^L = \vac{D^L(U_\mu^g)}$ and
$ Z_F^R = \vac{D^R(U_\mu^g)}$ at the leading order.
The wave function renormalizations for the bosonic fields are defined by
\eqs{
\sum_n < L^a_\mu(0) L^b_\nu(n) >
e^{ipa\cdot n} \nonumber
\quad \simeq \displaystyle \delta_{ab}
\frac{Z_A^L}{p^2 a^2 + m_G^2 a^2} \left
(\delta_{\mu\nu}+\frac{p_\mu p_\nu}{m_G^2} \right)
\label{lgauge}}
and
\eqs{
\sum_n < R^a_\mu(0) R^b_\nu(n) >
e^{ipa\cdot n} \nonumber
\quad \simeq \displaystyle \delta_{ab}
\frac{Z_A^R}{p^2 a^2 + m_G^2 a^2} \left
(\delta_{\mu\nu}+\frac{p_\mu p_\nu}{m_G^2} \right)  .
\label{rgauge}}
where $L_\mu^a = D^L ( A_\mu^a)$ and $R_\mu^a = D^R ( A_\mu^a)$.
{}From the consistency amomng $A_\mu$, $L_\mu$ and $R_\mu$, we obtain
\eqs{
\sum_n < L^a_\mu(0) A^b_\nu(n) >
e^{ipa\cdot n} \nonumber
\quad \simeq \displaystyle \delta_{ab}
\frac{\sqrt{Z_A^L\cdot Z_A}}{p^2 a^2 + m_G^2 a^2} \left
(\delta_{\mu\nu}+\frac{p_\mu p_\nu}{m_G^2} \right)
}
and
\eqs{
\sum_n < R^a_\mu(0) A^b_\nu(n) >
e^{ipa\cdot n} \nonumber
\quad \simeq \displaystyle \delta_{ab}
\frac{\sqrt{Z_A^R\cdot Z_A}}{p^2 a^2 + m_G^2 a^2} \left
(\delta_{\mu\nu}+\frac{p_\mu p_\nu}{m_G^2} \right)  .
}
The renormalized fermion gauge couplings $g_L^r$ and $g_R^r$ are both
non-zero and the left-handed coupling $g_L^r$ is different from the
right-handed coupling $g_R^r$ if $D^L \not= D^R$.
In the hypercharge sector of the standard model,
we expect that $g_L^r \propto Y_L$ and $g_R^r \propto Y_R$ in
the scaling region. Here $Y_L$ ( $Y_R$ ) is the hypercharge of the
left-handed ( right-handed ) fermion;
$ \displaystyle D^{L(R)}( U ) = U^{Y_{L(R)}}$.

Final remark is the following. The hopping parameter expansion we use
relates the fermionic quantities such as the
propagator or the 3-points function to the
bosonic quantities such as $L$, $Z_A$ or $Z_H$, which, however, can
not be calculated by this expansion.  We have to calculate these quantities
by other methods such as a mean-field calculation, a $1/d$ expansion or
numerical simulations.  In particular, a reliable evaluation for $Z_A$ is
crucial for the Wilson-Yukawa approach, as mentioned before.

\vspace{2cm}
\section*{Acknowledgements}
\bigskip
I would like to thank
Profs.  Y. Kikukawa,  H. Nielsen, J. Shigemitsu, R. Shrock and
A. Ukawa for useful discussions.

%\newpage

\end{document}